\documentclass[prl, twocolumn, superscriptaddress,nofootinbib,aps]{revtex4-2}
\usepackage{graphicx}
\usepackage{bm, amsmath, amssymb}
\usepackage{color}
\usepackage{soul}
\usepackage[T1]{fontenc}
\usepackage{amstext}
\usepackage{amsfonts}
\usepackage{float}
\usepackage{algorithm}
\usepackage{algpseudocode}
\usepackage{multirow}
\usepackage{physics}
\makeatletter
\def\frontmatter@thefootnote{%
 \altaffilletter@sw{\@fnsymbol}{\@fnsymbol}{\csname c@\@mpfn\endcsname}}%
\makeatother

\renewcommand{\eqref}[1]{\mbox{Eq.~(\ref{#1})}}

\newcommand{\figref}[1]{Fig.~\hyperref[#1]{\ref*{#1}}}
\newcommand{\figpanel}[2]{Fig.~\hyperref[#1]{\ref*{#1}(#2)}}
\newcommand{\figpanels}[3]{Fig.~\hyperref[#1]{\ref*{#1}(#2)-(#3)}}
\newcommand{\figpanelNoPrefix}[2]{\hyperref[#1]{\ref*{#1}(#2)}}

\usepackage[colorlinks=true,urlcolor=blue,citecolor=blue,linkcolor=blue]{hyperref}
\makeatother

\usepackage{babel}

\begin{document}
\title{Bridging Pulse and Circuit Levels: A Geometric Framework for Deterministic Error Suppression}
\author{Junkai Zeng}
\affiliation{International Quantum Academy, Shenzhen, 518048, China}
\author{Hao Liang}
\affiliation{International Quantum Academy, Shenzhen, 518048, China}
\author{Yong-Ju Hai}
\affiliation{International Quantum Academy, Shenzhen, 518048, China}
\author{Xiu-Hao Deng}
\email{dengxiuhao@iqasz.cn}
\affiliation{International Quantum Academy, Shenzhen, 518048, China}
\affiliation{Shenzhen Branch, Hefei National Laboratory, Shenzhen,518048, China.}

\begin{abstract}
Quantum errors in noisy environments remain a major obstacle to advancing quantum information technology. Standard quantum error correction requires massive ancillary qubit overhead, motivating the need for hardware-efficient mitigation strategies. In this work, we propose a framework for dynamical quantum error correction at the circuit level, requiring no logical encoding or ancillary qubits. By extending a geometric framework—originally developed for creating dynamical error-correcting gates at the control pulse level —to the discrete dynamics of digital circuits, we map the accumulation of coherent errors to trajectories in a high-dimensional error space. We demonstrate that inserting deterministically optimized twirling gate sequences actively shapes these trajectories, utilizing destructive interference to keep the accumulated error bounded with error scaling $\mathcal{O}(1)$. This deterministic path-shaping suppresses circuit errors fundamentally differently than the stochastic random-walk behavior of standard randomized compiling with error scaling $\mathcal{O}(\sqrt{N})$. Furthermore, we show that this circuit-level dynamical correction synergizes with pulse-level robust control, providing an analytical bridge between continuous noise dynamics and discrete quantum compilation. This research illuminates pathways to achieving highly noise-resistant quantum circuits prior to the era of fault tolerance.
\end{abstract}

\maketitle

Quantum technology holds the potential to outperform classical systems in critical computational tasks~\cite{arute2019quantum,wu2021strong,daley2022practical}. However, realizing this potential is fundamentally bottlenecked by environmental noise and imperfect control. While standard quantum error correction provides a theoretical pathway to fault tolerance, it demands massive ancillary qubit overhead and complex logical encoding, making it highly resource-intensive for near-term architectures. This reality has driven the pursuit of hardware-efficient error mitigation and robust control techniques that operate without the burden of logical qubit encoding. Extensive progress has been made at the continuous-time level, where robust pulse engineering maintains high-fidelity operations in noisy environments~\cite{bentleyNumericOptimizationConfigurable2020,codyjonesDynamicalDecouplingQubit2012,edmundsDynamicallyCorrectedGates2020,greenArbitraryQuantumControl2013,hansenPulseEngineeringGlobal2021,kabytayevRobustnessCompositePulses2014,khodjastehDesigningPracticalHighfidelity2013,khodjastehArbitrarilyAccurateDynamical2010,khodjastehDynamicallyErrorCorrectedGates2009,khodjastehFaultTolerantQuantumDynamical2005,sauvageOptimalControlFamilies2022}, significantly boosting gate fidelities across diverse physical platforms~\cite{veldhorst2015two,huang2019fidelity,xue2022quantum,mills2022two,sheldon2016procedure,krantz2019quantum,werninghaus2021leakage,egan2021fault,pogorelov2021compact}.

A vital, complementary frontier is embedding dynamical error correction directly at the discrete circuit level, independent of underlying pulse implementations. Existing circuit-level strategies include dynamical decoupling for resting qubits~\cite{khodjastehArbitrarilyAccurateDynamical2010,khodjastehFaultTolerantQuantumDynamical2005}, constructing dynamically error-corrected gates via sequences of trivial gates~\cite{khodjastehDynamicallyErrorCorrectedGates2009}, hidden inverses for coherent error cancellation~\cite{zhang2022hidden}, crosstalk-aware compilation~\cite{xie2022suppressing}, and randomized compiling (RC)~\cite{wallmanNoiseTailoringScalable2016,hashimRandomizedCompilingScalable2021,guNoiseresilientPhaseEstimation2022,urbanek2021mitigating,perrin2024mitigating}. While standard RC successfully mitigates coherent errors, it relies heavily on the statistical averaging of an ensemble of varying circuits to stochasticize noise, which inherently requires execution overhead and does not deterministically eliminate the error of a single execution shot.

\begin{figure}[tbp]
\includegraphics[width=\columnwidth]{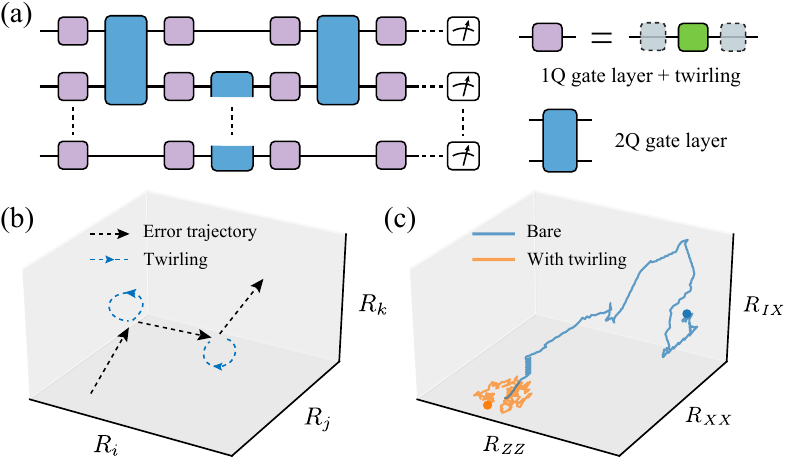} 
\caption{A conceptual overview of dynamical quantum error correction at the circuit level via deterministic geometric twirling. (a) A generic circuit structure modified by inserting specifically chosen twirling gates. (b) Conceptual illustration of error accumulation as a trajectory in the quantum erroneous evolution diagram~\cite{hai2025universal}. A standard noisy gate adds a step vector to the error path, while an ideal twirling gate rotates the reference frame, altering the direction of subsequent error steps. (c) Comparison of error accumulation strategies through illustrative two-qubit circuit example: the unmitigated bare circuit allows errors to accumulate constructively (orange path); standard randomized compiling converts the trajectory into a stochastic random walk (blue path); whereas deterministic geometric twirling actively bounds the error trajectory near the origin via engineered destructive interference. Although the true error space is high-dimensional, a 3D subspace is shown for intuitive visualization.}
\label{fig:RC_picture}
\end{figure}

In this work, we introduce a framework for dynamical quantum error correction at the circuit level that deterministically suppresses coherent errors without encoding logical qubits or utilizing ancillary qubits. We achieve this by extending a geometric framework—originally developed for continuous-time pulse control~\cite{zeng2018general,zeng2018fastest,zhuangNoiseresistantLandauZenerSweeps2022,dongDoublyGeometricQuantum2021,buterakosGeometricalFormalismDynamically2021,barnesDynamicallyCorrectedGates2022, hai2025universal}—to the discrete dynamics of gate-based circuits. This geometric framework visualizes the accumulation of coherent error as a trajectory in an abstract, high-dimensional error space, forming a diagram called the quantum erroneous evolution diagram (QEED) \cite{hai2025universal, yi2024robust}. By conceptualizing error accumulation geometrically, we demonstrate that constructing a robust circuit is equivalent to deterministically shaping this error trajectory to remain bounded near the origin~\cite{hai2025universal}. Rather than relying solely on the statistical "random walk" of RC~\cite{wallmanNoiseTailoringScalable2016}, our framework enables the strategic insertion of optimized twirling sequences that induce destructive interference of errors across circuit layers. Fig.~\ref{fig:RC_picture} provides a conceptual illustration of this deterministic geometric twirling. We will formalize this geometric approach, show how it analytically bridges discrete circuit compilation with continuous noise dynamics, and demonstrate that circuit-level dynamical correction synergizes powerfully with pulse-level robust control~\cite{hai2025universal}.

\paragraph*{Geometric error analysis at the circuit level.-}Here, we introduce our geometric formalism by first studying generic noisy quantum dynamics and then extending our analysis to discrete circuit models. Consider a multi-qubit system subject to coherent noise, described by the Hamiltonian: $H(t)=H_{0}(t)+\sum_{i}\epsilon _{i}(t)V_{i}$. Here $H_{0}$ is the noise-free part of the Hamiltonian, $V_{i}$ the noise operator, and $\epsilon _{i}(t)$ the noise strength. The time dependency of the noise $\epsilon _{i}(t)$ can take various forms, such as constant, randomly fluctuating, drifting, or proportional to the control amplitude, depending on the noise source. For coherent noise, $V_{i}$ is typically the tensor product of Pauli matrices for qubit systems.

The evolution operator under weak noise can be decomposed into two parts: 
\begin{equation}
U(t)=U_0(t)U_{\epsilon }(t),  
\label{eq:decom}
\end{equation}%
where the error-free evolution $U_{0}(t)=\mathcal{T}e^{-i\int dsH_{o}(s)}$ and the noise-induced deviation $U_{\epsilon }(t)=\mathcal{T}e^{-i\Phi (t)}$, with $\Phi (t)=\int_{0}^{t}ds\sum_{i}\epsilon _{i}(s)U_{0}^{\dagger}(s)V_{i}U_{0}(s)$ referred to as the \textit{error operator} in the following. As the noise $\epsilon _{i}(t)$ is generally weak, we apply the Magnus expansion and truncate higher-order terms so that $U_{\epsilon }(t)\approx e^{-i\Phi (t)}$ holds. Expanding the $n$-qubit error operator in the Pauli basis, $\sigma_{j}\in span\{I,X,Y,Z\}^{\otimes n}$, we write
\begin{equation}
\Phi (t)=\sum_{ij}\sigma_{j}\int_{0}^{t}\epsilon
_{i}(s)T_{j}^{(i)}(s)ds\equiv \sum_{ij}\sigma_{j}\mathbf{R}%
_{j}^{(i)}(t)  \label{eq:timedependent}
\end{equation}%
Here $i=1,...,N_{\mathrm{src}}$ labels all independent noise sources and $j=1,...,4^{n}-1$ labels the non-identity $n$-qubit Pauli basis elements. $T_{j}^{(i)}(t)=\frac{1}{D}\Tr\left( \sigma_{j}U_{0}^{\dagger }(t)V_{i}U_{0}(t)\right) $ represents the first-order instantaneous susceptibility of the operator $\sigma_{j}$ to the noise component $V_{i}$ at time $t$, and $D$ is the rank of $V_i$. The vector function $\mathbf{R}^{(i)}(t)$ characterizes the amount of error accumulated on each operator basis $\sigma_{j}$ at time $t$. Geometrically, the point $\mathbf{R}^{(i)}(t)$ traces a trajectory in the $(4^{n}-1)$-dimensional space spanned by all Pauli operators $\sigma_{j}$, except the identity $I$. For normalized $V_{i}$, this point moves with velocity $\epsilon _{i}(t)$ in the direction $\mathbf{T}^{(i)}(t)$. The trajectory forms a geometric path whose properties encode the error accumulation dynamics. This geometric framework, visually represented as QEED~\cite{hai2025universal}, provides an intuitive illustration of how errors build up over time.

Crucially, the error operator $\Phi (t)$ depends not only on the noise $V_{i}$ but also on the interaction picture transformation through $U_{0}(t)$. This geometric insight demonstrates that different control strategies produce distinct error trajectories even under identical noise conditions. In the continuous-time regime, this has been exploited to design robust control waveforms that create closed-loop trajectories in the QEED, effectively canceling errors~\cite{hai2025universal}.

Here, we generalize the geometric QEED approach from continuous dynamics to discrete quantum circuits, where the continuous evolution is discretized into circuit layers: $\sum_{i}R_{j}^{(i)}(t)\rightarrow \Gamma_{j}^{[l]}$. Here, $\mathbf{\Gamma}^{[l]}$ represents the error vector step contributed by layer $l$. We omit the noise source index $i$ by absorbing the summation over noise sources into the definition of $\Gamma$. 

To formalize this for discrete circuits, we distinguish between local errors and their global contributions. Thus, the error step vector after this move $\mathbf{\Gamma}^{[l]}$ is connected with local error steps $\mathbf{\Gamma}_{\text{lo}}^{[l]}$ by $\Gamma _{j}^{[l]}=\frac{1}{D}\Tr(\mathcal{C}^{[l]\dagger }\left( \mathbf{\Gamma }_{\text{lo}}^{[l]}\cdot \mathbf{\sigma }\right) \mathcal{C}^{[l]}\sigma_{j})$, which leads to the total accumulated error operator: 
\begin{equation}
\Phi_{\text{total}}=\sum_{l=1}^{N_{D}}\sum_{j}\Gamma _{j}^{[l]}\sigma_{j}  
\label{eq:Phitotal}
\end{equation}
Here $N_{D}$ is the circuit depth, $j$ covers all indices of the generalized Pauli matrices, and $\mathcal{C}^{[l]}$ represents the circuit from the 1st to $l$'th layer. Since $\mathcal{C}^{[l]}$ is unitary and thus norm-preserving, error vectors $\mathbf{\Gamma }^{[l]}$ and $\mathbf{\Gamma }_{\text{lo}}^{[l]}$ are connected through a rotation transformation. Eq.~\ref{eq:Phitotal} can be seen as a path formed by $N_{D}$ steps in the QEED, with each step vector being $\mathbf{\Gamma} ^{[l]}$.

\paragraph*{Deterministic Bounding of Coherent Error Trajectories in Quantum Circuits.-}
Building upon this geometric formalism, we now demonstrate how to deterministically suppress coherent errors by optimizing circuit robustness with specific twirling sequences. A quantum circuit can be structurally decomposed into a series of alternating "easy" and "hard" layers (where consecutive easy layers are combined, and ideal identity layers are inserted between consecutive hard layers), as depicted by the purple and blue boxes in Fig.~\ref{fig:RC_picture}(a). To generate an equivalent but dynamically protected circuit, we insert a twirling operator $T$ before a hard layer $l$ and a corresponding correction operator $T^c$ afterward. This ensures the modified layer performs the exact same logical operation, satisfying $T^{c}U_{0}^{[l]}T=U_{0}^{[l]}$. If the twirling operators are restricted to the Pauli group and the hard layer consists of Clifford gates, the correction operators $T^c$ are identically Pauli gates and can be efficiently calculated. Under the easy-hard decomposition, these twirling and correction operators can be absorbed directly into the adjacent single-qubit gates, including the identity operations between consecutive hard layers, ensuring that this deterministic protection adds no hard layer overhead to the quantum circuit at logical level. At the hardware level, however, we shall mention that the modified single-qubit layers can introduce additional gate-time duration, recalibration, and implementation costs depending on hardware architectures.

The twirled noisy hard layer is represented as $T^{c}U_{0}^{[l]}e^{-i\left( \mathbf{\Gamma }_{\text{lo}}^{[l]}\cdot \mathbf{\sigma }\right) }T$. By transforming the error operator through the twirling gate, we obtain $U_{0}^{[l]}e^{-iT^{\dagger }\left( \mathbf{\Gamma }_{\text{lo}}^{[l]}\cdot \mathbf{\sigma }\right) T} \equiv U_{0}^{[l]}e^{-i \tilde{\mathbf{\Gamma }}^{[l]}_{\text{lo}}\cdot \mathbf{\sigma }}$. Geometrically, this action corresponds to the twirling operation $T$ rotating the local step error vector $\mathbf{\Gamma }_{\text{lo}}^{[l]}$ into a new orientation $\tilde{\mathbf{\Gamma }}_{\text{lo}}^{[l]}$. This process actively modifies the error trajectory in the QEED associated with the circuit, as illustrated by the dashed blue curves in Fig.~\ref{fig:RC_picture}(b). As the circuit depth scales to $N_{D}$ layers, the total error distance of the deterministically twirled circuit becomes $\left|\sum_{i}^{N_{D}}\mathbf{\tilde{\Gamma}} ^{[i]}\right|$, directly contrasting with the unmitigated original circuit's error distance $\left|\sum_{i}^{N_{D}}\mathbf{\Gamma} ^{[i]}\right|$. Here, the global error step $\mathbf{\tilde{\Gamma}} ^{[i]}$ is obtained from the local twirled error $\mathbf{\tilde{\Gamma}}_{\text{lo}} ^{[i]}$ via the same reference frame transformation as the original circuit: $\tilde{\Gamma} _{j}^{[i]}=\frac{1}{D}\Tr(\mathcal{C}^{[i]\dagger }\left( \tilde{\mathbf{\Gamma }}_{\text{lo}}^{[i]}\cdot \mathbf{\sigma }\right) \mathcal{C}^{[i]}\sigma_{j})$. 

Instead of relying on statistical random walks, we can deterministically design a specific sequence of twirling gates to explicitly satisfy $\left|\sum_{i}^{N_{D}}\mathbf{\tilde{\Gamma}} ^{[i]}\right|<\left|\sum_{i}^{N_{D}}\mathbf{\Gamma} ^{[i]}\right|$, thereby guaranteeing error suppression for a single shot. Fig.~\ref{fig:RC_picture}(c) visually illustrates how an optimized sequence of Pauli twirling gates forces the error trajectory to remain bounded near the origin. While randomly chosen twirling sequences might inadvertently align error vectors and increase total error, an optimized deterministic sequence acts as a circuit-level dynamical decoupling protocol. Finding the global optimum for these sequences is generally computationally hard; however, various heuristic optimization methods can be employed. For shallow circuits, brute-force searching is highly effective, and a reduced set of twirling gates can be utilized when the target noise channel does not span the entire Pauli space~\cite{cai2019constructing}. 

\begin{figure}[tbp] 
\includegraphics[width=\columnwidth]{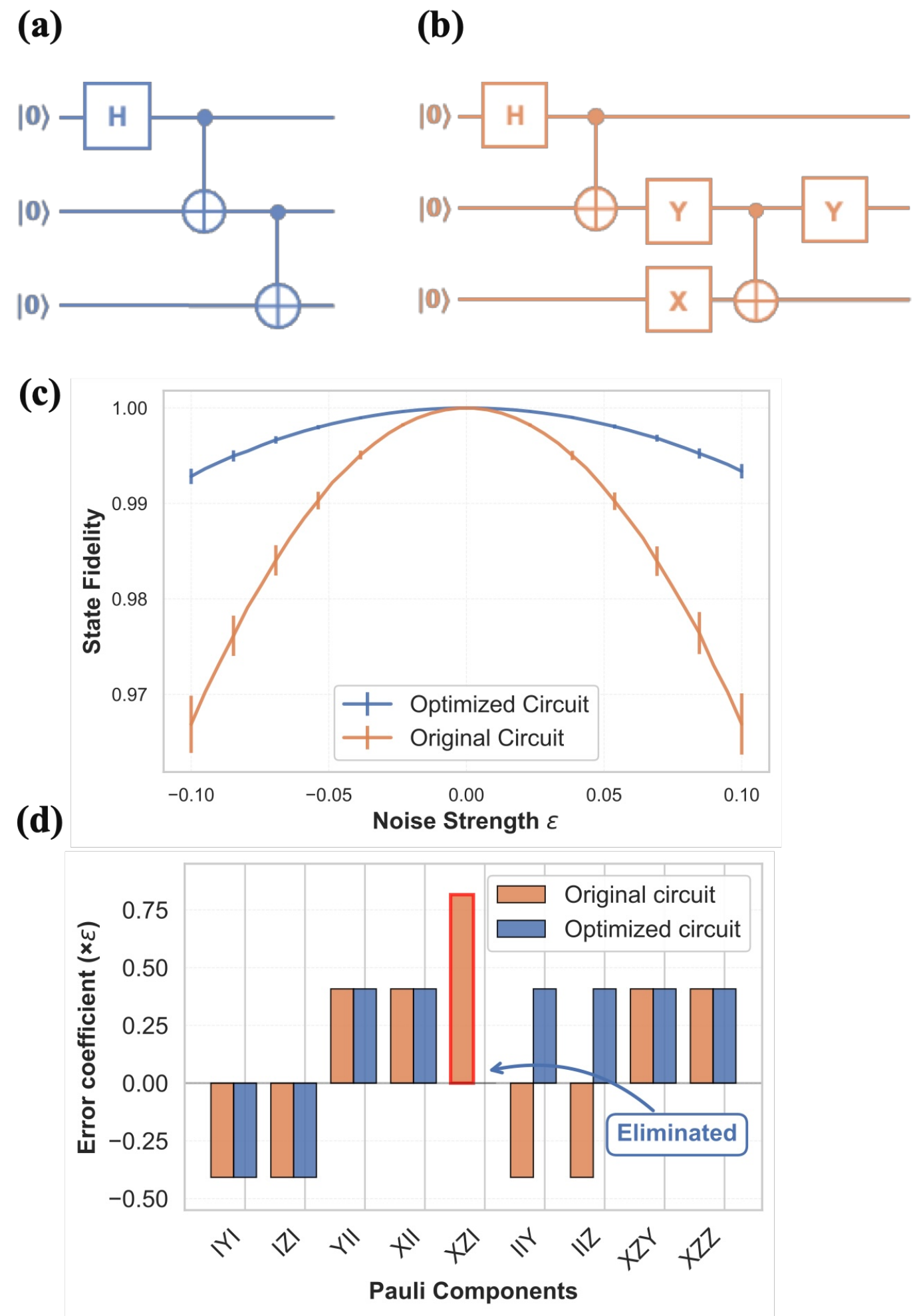}
\caption{Optimization of a 3-qubit GHZ state preparation circuit subjected to the coherent gate error $(Z+Y)I+(Z-I)(Y+Z)$ on imperfect CNOT gates. (a) Standard unmitigated circuit. (b) Equivalent circuit protected by deterministic geometric twirling gates, where the single-qubit operations replacing identity between consecutive CNOT layers are explicitly displayed. (c) A comparison of state fidelity versus noise strength. The error bars represent standard deviations over the sampled perturbations. (d) Pauli decomposition of the accumulated error operator. The bare circuit develops a dominant $XZI$ component (highlighted in red) due to constructive interference between CNOT layers. The optimized deterministic twirling sequence (blue bars) intentionally induces destructive interference, completely eliminating this dominant term.}
\label{fig:GHZ_circuit}
\end{figure}

To concretely demonstrate this deterministic path-shaping, we apply our framework to a Greenberger–Horne–Zeilinger (GHZ) state preparation circuit, as depicted in Fig.~\ref{fig:GHZ_circuit}(a) and (b). In this model, the CNOT gates are afflicted by the coherent error operator $\Phi =\epsilon \frac{1}{\sqrt{6}}\left((Z+Y)I+(Z-I)(Y+Z)\right)$. We selected this specific error model to represent a realistic combination of local coherent errors (e.g., $ZI$, $YI$) and correlated two-qubit errors (e.g., $ZY$, $ZZ$) arising from phenomena such as control crosstalk or residual ZZ coupling. The equivalent robust circuit shown in Fig.~\ref{fig:GHZ_circuit}(b) is generated by strategically inserting Pauli twirling gates (and their corrections) around the CNOT operations. By absorbing these twirling operations into the adjacent easy single-qubit layers, we alter the physical single-qubit gates while preserving the exact overall logic of the GHZ circuit. Our numerical analysis confirms that this deterministically twirled circuit exhibits significantly higher resilience against coherent noise compared to the bare circuit, as validated by the fidelity scaling in Fig.~\ref{fig:GHZ_circuit}(c). 
We further assess the sensitivity of the protection to imperfect noise characterization by evaluate the circuit under perturbed coherent error models. Each term in the coherent error operator,
$\Phi=\epsilon \sum_{\alpha} c_{\alpha} P_{\alpha}$, is randomly deviated such that  $\Phi_{\mathrm{pert}}=\epsilon \sum_{\alpha} c_{\alpha}(1+\delta_{\alpha}) P_{\alpha}$,
where each coefficient is independently varied by $\delta_{\alpha}\sim \mathrm{Unif}[-0.2,\,0.2]$. For each value of the noise strength $\epsilon$, we sample multiple perturbed error instances and compute the corresponding output-state fidelities. The smaller error bars on the fidelity curve of the twirled circuit in Fig.~\ref{fig:GHZ_circuit}(c) indicates that the deterministic protection remains effective under moderate noise model mismatch.

\paragraph*{The statistical limit: Randomized compiling as a geometric random walk.-}While deterministic geometric twirling successfully bounds the error trajectory near the origin for a single shot (achieving $\mathcal{O}(1)$ error scaling), finding the global optimum for deep circuits can be computationally intensive. It is physically illuminating to examine the ensemble average of all possible twirled trajectories. When twirling gates are sampled randomly rather than being deterministically optimized, the engineered destructive interference is lost. Instead, the error step vectors randomly flip signs, and the trajectory inevitably degrades into a random walk in the generalized Pauli space. This statistical limiting case of our geometric framework corresponds directly to standard RC~\cite{wallmanNoiseTailoringScalable2016}.

In this RC limit, multiple instances of randomly twirled circuits are averaged classically. We now analyze the resulting mitigation effects, focusing on the behavior of a single circuit layer. The noise channel on operator $P$ for one circuit layer is $\mathcal{E}(P) = \mathbb{E}\left(\tilde{U}_\epsilon P \tilde{U}_\epsilon^\dagger\right)$, where the expectation operator $\mathbb{E}$ averages over all possible twirling gates. Because this averaging produces a mixed (non-unitary) evolution, the resulting noise channel is a superoperator $\mathcal{E}(\cdot)$. We rewrite the error of a layer using the Magnus expansion on the noise strength $\epsilon$, $\tilde{U}_\epsilon = \exp\left(-i\sum_{n=1}^{\infty} \frac{1}{n!}\sum_i \epsilon^n \Gamma_i^{(n)}T\sigma_i T\right)$, where the $n$-th order error term is measured by $\Gamma_i^{(n)}$ and $T$ is the twirling operator. (Note: the parenthesized upper index of $\Gamma_i^{(n)}$ indicates the order in the Magnus expansion, not the circuit layer). Performing a Taylor expansion on $\mathcal{E}(P)$ around $\epsilon=0$ and averaging over $T$ yields the analytical form of the resulting Pauli channel. The first four orders, $\mathcal{E}(P)=\mathcal{E}_\epsilon^{0-3}(P)+\mathcal{E}_\epsilon^{4}(P)$, are calculated as:
\begin{equation}
\label{eq:noisechannelexp}
    \begin{split}
    &\mathcal{E}_\epsilon^{0-3}(P) = P - 2\sum_{\{\sigma_i,P\}=0} \Gamma_{i}^{(1)}\left( \Gamma^{(1)}_i +\Gamma^{(2)}_i\right) P \\
        &\mathcal{E}_\epsilon^{4}(P)=\sum_{\{\sigma_{i},P\}=0} \left( \frac{2}{3}\Gamma^{(1)}_i \left(\left(\Gamma^{(1)}_i\right)^3-\Gamma^{(3)}_i\right)-\frac{1}{2}\left(\Gamma^{(2)}_i\right)^2 \right)P\\ &+\sum_{\substack{\{\sigma_{\{i,j\}},P\}=0\\
                                [\sigma_i, \sigma_j]=0\\
                                i\neq j
                                }
                                    }2\left(\Gamma^{(1)}_i \Gamma^{(1)}_j\right)^2 P
    \end{split}
\end{equation}
Here, the coherent cross-terms of the twirling gates have been integrated out by the averaging process (detailed derivations are provided in Sect.~\ref{noisech} of the supplementary material). 

Deriving this analytical form directly links the phenomenological Pauli noise channel back to the geometric error distance of individual gates ($\Gamma_i^{(n)}$), providing a vital tool for error assessment and post-processing mitigation.

\begin{figure}[tbp]
    \includegraphics[width=\columnwidth]{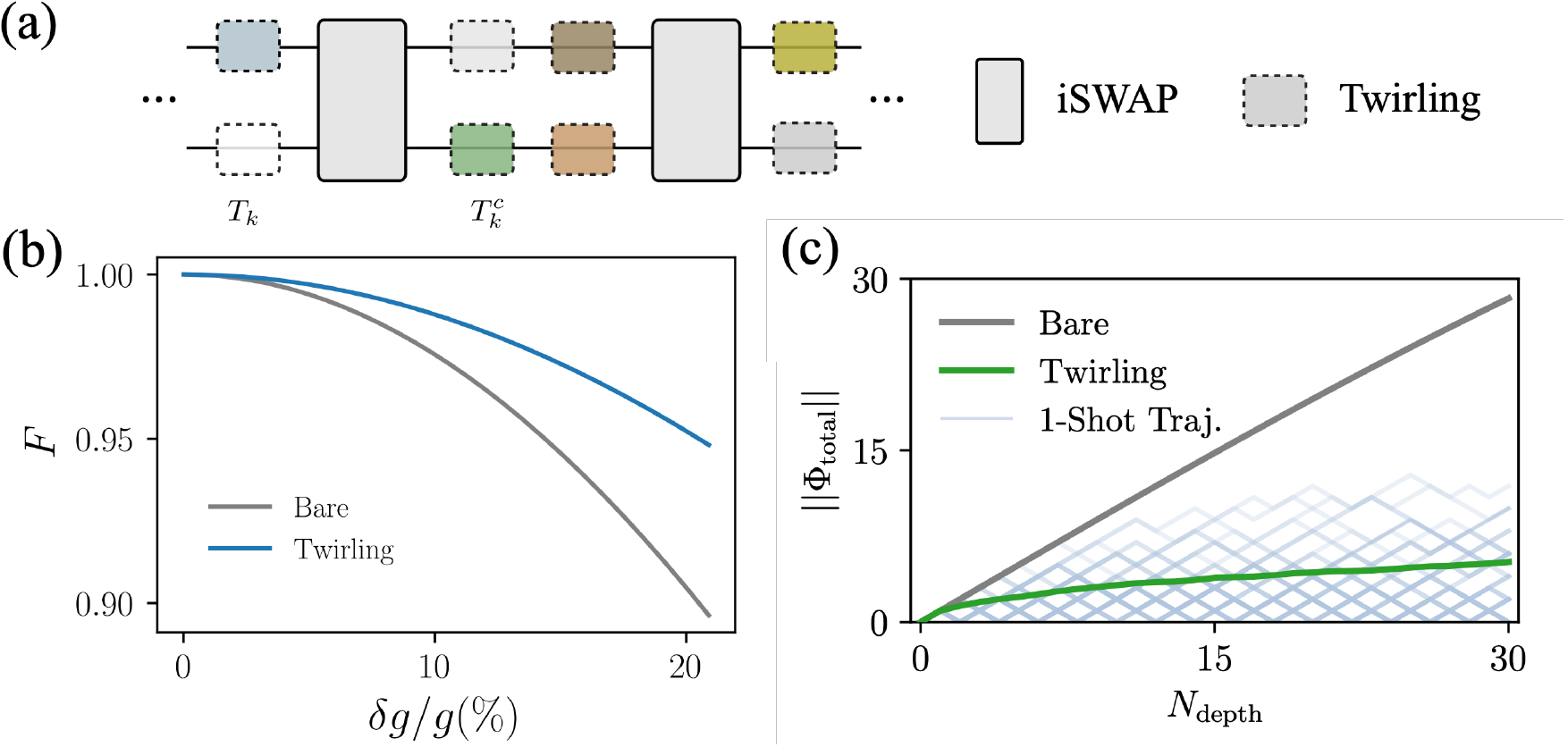}
    \caption{(a) Visualization of an iSWAP circuit with random twirling gates added. (b) State transfer fidelity of a single layer. (c) Error distances as circuit depth grows. The single-shot trajectory ("1-Shot Traj.") varies wildly, underscoring the risk of unoptimized sequences, while the RC ensemble average ("Twirling") clearly exhibits $\mathcal{O}(\sqrt{N})$ random walk scaling, compared to the unmitigated $\mathcal{O}(N)$ linear accumulation.}
    \label{fig:iSWAP}
\end{figure}

The profound difference in error scaling is clearly observable in the iSWAP circuit model (Fig.~\ref{fig:iSWAP}). An unmitigated circuit accumulates error linearly, leading to $\mathcal{O}(N)$ scaling. As shown in Fig.~\ref{fig:iSWAP}(c), applying random twirling causes a single execution shot ("1-Shot Traj.") to vary wildly—which is precisely why the deterministic selection discussed in Sec. III is vastly superior for single-shot execution. However, when multiple random instances are averaged via RC, the mean error distance increases proportionally to the square root of the circuit depth ($\mathcal{O}(\sqrt{N})$), perfectly consistent with an $N$-dimensional random walk. 

\begin{figure*}[tbp]
    \includegraphics[width=2\columnwidth]{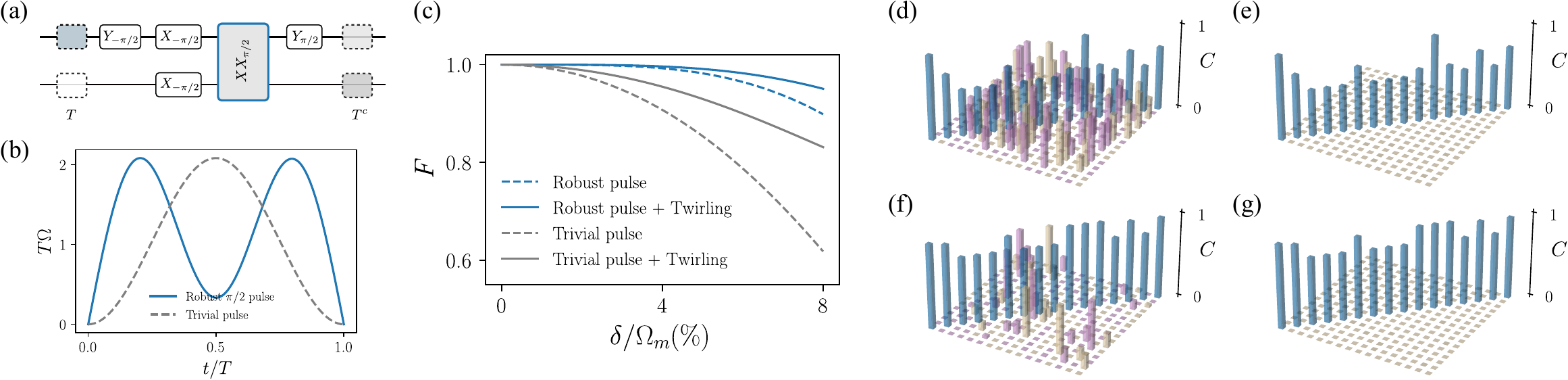}
    \caption{(a) Visualization of the CNOT circuit with dotted line boxes representing random twirling gates. (b) Dephasing-robust optimized control pulse (blue) versus a trivial cosine pulse (grey). (c) State transfer fidelity comparison. (d)-(g) Pauli Transfer Matrices (PTMs) of the noise channel under varying conditions. Diagonal elements are blue; negative and non-negative off-diagonal elements are plum and wheat. The robust pulse (f) maximizes the diagonals by shrinking $\Gamma^{(1)}$, while RC (e) purifies the off-diagonals. The combined approach (g) synergizes both.} 
    \label{fig:CNOT}
\end{figure*}

Crucially, Eq.~\ref{eq:noisechannelexp} reveals a powerful synergy between continuous pulse-level robustness and discrete circuit-level compiling. While RC purifies the noise channel by averaging out coherent off-diagonal errors, the fundamental severity of the channel is still dictated by the gate's inherent error distance $\Gamma_i^{(1)}$. By utilizing pulse-level robust control to engineer closed curves in the continuous QEED, we can force $\Gamma_i^{(1)}=0$. As dictated by Eq.~\ref{eq:noisechannelexp}, this eliminates the dominant first-order decay terms entirely.

We numerically demonstrate this synergy using a composed CNOT circuit subjected to dephasing noise (Fig.~\ref{fig:CNOT}). The Pauli Transfer Matrices (PTMs) visually capture the distinct and complementary benefits. Robust pulse control (Fig.~\ref{fig:CNOT}f) directly shrinks the fundamental error step size, pushing the diagonal elements closer to unity. RC (Fig.~\ref{fig:CNOT}e) acts purely as a spatial averager, eliminating off-diagonal coherent cross-terms. Combining both techniques (Fig.~\ref{fig:CNOT}g) provides the ultimate noise resilience: robust control minimizes the incoherent error rates at the source, and RC purifies the remainder into a highly mitigated, stochastic Pauli channel. This confirms that gate-level continuous control and circuit-level discrete compiling are fundamentally complementary within our geometric framework.

\paragraph*{Discussion and Conclusion.-}In this work, we have established a geometric framework for realizing circuit-level dynamical quantum error correction without the massive overhead of ancillary logical encoding. By mapping the discrete accumulation of coherent circuit errors to trajectories within the quantum erroneous evolution diagram, we demonstrated that noise resilience can be deterministically engineered. Rather than passively relying on the $\mathcal{O}(\sqrt{N})$ stochastic random walk of standard randomized compiling, our approach utilizes deterministic geometric twirling to actively induce destructive interference, bounding the error trajectory near the origin. Furthermore, our analytical derivation of the effective noise channel directly connects discrete compilation parameters to the continuous error distance of individual gates. This reveals a powerful synergy: robust continuous-time quantum control minimizes the fundamental gate error bounds, while optimal circuit-level twirling eliminates the residual coherent accumulation, providing a comprehensive, cross-stack mitigation strategy.

To cleanly demonstrate these geometric principles, our initial analysis approximated single-qubit gates as ideal. This is physically justified in current architectures, where typical single-qubit fidelities ($>99.9\%$) surpass two-qubit gate fidelities ($\sim 99\%$) by an order of magnitude. However, the framework is intrinsically extensible to deeper circuits or platforms with comparable error rates; single-qubit layers simply contribute their own localized error vectors, $\mathbf{\Gamma}_{\text{single}}^{[l]}$, to the global trajectory. Crucially, absorbing twirling and correction operators directly into these single-qubit layers ensures that robustness is enhanced without increasing overall circuit depth.

The primary bottleneck in extending this deterministic path-shaping to arbitrary, large-scale quantum circuits is the discrete optimization of twirling sequences, a problem that scales exponentially with circuit size. Overcoming this combinatorial complexity defines the next critical frontier in quantum compilation. For highly structured and repetitive applications, such as variational quantum eigensolvers, block-wise optimization can be effectively leveraged and reused. For general, unstructured circuits, deploying advanced heuristic solvers—including deep reinforcement learning, genetic algorithms, and greedy layer-by-layer optimization—will be essential. Solving this optimization challenge will fully unlock the power of zero-overhead, deterministic geometric error suppression in the pre-fault-tolerant era.

\begin{acknowledgments}
This work was supported by the Key-Area Research and Development Program of Guang-Dong Province (Grant No. 2018B030326001), the National Natural Science Foundation of China (U1801661), the Guangdong Innovative and Entrepreneurial Research Team Program (2016ZT06D348), the Guangdong Provincial Key Laboratory (Grant No.2019B121203002), the Natural Science Foundation of Guangdong Province (2017B030308003), and the Science, Technology and Innovation Commission of Shenzhen Municipality (JCYJ20170412152620376, KYTDPT20181011104202253), and the NSF of Beijing (Grants No. Z190012), Shenzhen Science and Technology Program (KQTD20200820113010023), and the National Natural Science Foundation of China (Grant No 12404566)  
\end{acknowledgments}

\bibliography{apssamp}

\clearpage
\onecolumngrid

\begin{center}
\textbf{\large Supplementary Materials for: Bridging Pulse and Circuit Levels: A Geometric Framework for Deterministic Error Suppression} \\[0.5cm]
Junkai Zeng,$^{1}$ Hao Liang,$^{1}$ Yong-Ju Hai,$^{1}$ and Xiu-Hao Deng$^{1, 2, *}$ \\[0.2cm]
{\small \it $^{1}$International Quantum Academy, Shenzhen, 518048, China} \\
{\small \it $^{2}$Shenzhen Branch, Hefei National Laboratory, Shenzhen, 518048, China} \\[0.1cm]
{\small \it $^*$Email: dengxiuhao@iqasz.cn}
\end{center}
\vspace{0.5cm}

\setcounter{secnumdepth}{3}
\setcounter{page}{1}
\renewcommand{\thepage}{S\arabic{page}}
\setcounter{section}{0}
\renewcommand{\thesection}{S\arabic{section}}
\setcounter{equation}{0}
\renewcommand{\theequation}{S\arabic{equation}}
\setcounter{figure}{0}
\renewcommand{\thefigure}{S\arabic{figure}}
\setcounter{table}{0}
\renewcommand{\thetable}{S\arabic{table}}

\section{Distribution of Randomly Twirled Error Operator}
\label{App_Twirl}

We analyze the distribution of the randomly twirled first-order error using the Hilbert–Schmidt (HS) inner product. We consider only Pauli twirling, as more complex twirling would require correspondingly complex correction operators that are difficult to implement. We expand the local error at layer $i$ as $\mathbf{\Gamma}_{\text{lo}}^{[i]}=\sum_j \Gamma_{\text{lo},j}^{[i]}\sigma_j$ and the transported Pauli operators as $\mathcal{C}^{[i]\dagger}\sigma_j \mathcal{C}^{[i]}=\sum_k \mathcal{R}_{kj}^{[i]}\sigma_k$, where $\mathcal{R}^{[i]}$ is the Pauli transfer matrix representation of the unitary $\mathcal{C}^{[i]}$ and is real orthogonal. The total twirled error operator can be expressed as:
\begin{equation}
    \begin{split}
        \tilde{\Phi}_{\text{total}} &= \sum_{i=0}^{N_D} \sum_j \gamma^{[i]}_j \Gamma_{\text{lo},j}^{[i]}\left(\mathcal{C}^{[i]\dagger}\sigma_j\mathcal{C}^{[i]}\right)\\ 
&= \sum_{i=0}^{N_D} \sum_{j,k} \gamma^{[i]}_j \Gamma_{\text{lo},j}^{[i]}\mathcal{R}^{[i]}_{kj} \sigma_k\equiv\sum_k \tilde{\Phi}_k\sigma_k
    \end{split}
\end{equation}
where each $\gamma_j^{[i]}\in\{\pm1\}$ denotes the commutation sign between the twirling operator at layer~$i$ and~$\sigma_j$ ($+1$ if they commute, $-1$ if they anti-commute). When all single-qubit Pauli's are chosen with equal probability, $\mathbb{E}[\gamma_j^{[i]}]=0$ and the twirls on different layers are statistically independent, so that $\mathbb{E}[\tilde{\Phi}_{\mathrm{total}}]=0$.

For the second moment (Hilbert-Schmidt norm expectation), we have 

\begin{equation}
\begin{split}
    \mathbb{E} \|\tilde{\Phi}_{\text{total}}\|_{\text{HS}}^2 
    &= \sum_k \mathbb{E}(\tilde{\Phi}_k^2) \\ 
    &= \sum_k \sum_{i=0}^{N_D}\sum_j \left(\Gamma_{\text{lo},j}^{[i]}\right)^2\left(\mathcal{R}^{[i]}_{kj} \right)^2 \\
    & = \sum_{i=0}^{N_D}\sum_j \left(\Gamma_{\text{lo},j}^{[i]}\right)^2 \sum_k\left(\mathcal{R}_{kj}^{[i]}\right)^2\\
    &=\sum_{i=0}^{N_D} \|\mathbf{\Gamma}^{[i]}_{\text{lo}}\|_{\text{HS}}^2
\end{split}
\end{equation}
Here we used the orthogonality of $\mathcal{R}^{[i]}$ that $\sum_k \left(\mathcal{R}_{kj}^{[i]}\right)^2=1$.  Hence, if all layers have the same HS magnitude $\|\mathbf{\Gamma}_{\mathrm{lo}}^{[i]}\|_{\mathrm{HS}}=\Gamma$,
\begin{equation}
\sqrt{\mathbb{E}\|\tilde{\Phi}_{\mathrm{total}}\|_{\mathrm{HS}}^2}
   = \sqrt{N_D}\,\Gamma .
\end{equation}
By Jensen's inequality, $\mathbb{E}\|\tilde{\Phi}_{\mathrm{total}}\|_{\mathrm{HS}}
 \le \sqrt{\mathbb{E}\|\tilde{\Phi}_{\mathrm{total}}\|_{\mathrm{HS}}^2}
 = \sqrt{N_D}\Gamma$, with equality asymptotically in fully mixing (scrambling) regimes.

Each Pauli coefficient of the total error takes the form $\tilde{\Phi}_k=\sum_i a_{i,k}\gamma_i$, a sum of independent mean-zero $\pm1$ variables. By the Lindeberg central-limit theorem, $\tilde{\Phi}_k$ approaches a normal distribution with variance $\|\Gamma_{\mathrm{lo}}^{[i]}\|_{\mathrm{HS}}^2$.
Consequently $\|\tilde{\Phi}_{\mathrm{total}}\|_{\mathrm{HS}}$ concentrates around $\sqrt{N_D}\Gamma$ for large circuit depth $N_D$.

For comparison, if the random signs are fixed as $\gamma_j^{[i]}=1$ (no twirling), the total error is
\begin{equation}
\Phi
 = \sum_i \mathcal{C}^{[i]\dagger}\Gamma_{\mathrm{lo}}^{[i]}\mathcal{C}^{[i]},
\end{equation}
and
\begin{equation}
\|\Phi\|_{\mathrm{HS}}^2
 = \sum_i\|\Gamma_{\mathrm{lo}}^{[i]}\|_{\mathrm{HS}}^2
 + 2\!\!\sum_{i<j}\!\!
   \big\langle \Gamma_{\mathrm{lo}}^{[i]},
   \mathcal{C}^{[i]\dagger}\mathcal{C}^{[j]}
   \Gamma_{\mathrm{lo}}^{[j]}
   \mathcal{C}^{[j]\dagger}\mathcal{C}^{[i]}\big\rangle_{\!\mathrm{HS}} .
\end{equation}
The second term represents cross-correlations between layers.
Without twirling these correlations persist, but randomized compiling introduces statistically independent signs $\gamma^{[i]}_j$ that eliminate the cross terms in expectation, yielding the RMS scaling above.


\section{Computing Error Coefficients $\Gamma^{(n)}$}
\label{ErrorGamma}
In the main text we did not explicitly examine the detailed structure of the error model for clarity. 
Here we provide the general expressions for computing the error coefficients $\Gamma^{(n)}$, which fully characterize the perturbative noise channel.
For a general time-dependent noise, the first order error in the Magnus expansion of the error operator is $\Gamma^{(1)}_j=\sum_i R_j^{\epsilon(i)}(T)=\sum_i \int_0^T \epsilon_i(t) r_{i,j}'(t)dt$. Here we use $\Gamma_j^{(n)}$ to represent $n$'th order Magnus expansion for a circuit layer, and use $R_j^{\epsilon(i)}(t)$ for the corresponding continuous dynamics of the gate implementation. 
When the error coefficient $\Gamma_j^{(1)}$ appears in the noise channel after performing randomized compiling, it only appears in the second order term as its variance $\langle \left(\Gamma_j^{(1)}\right)^2\rangle$, which is given by 
\begin{equation}
\label{eq:channelavg}
\begin{split}
    \langle R_{j}^{\epsilon(i)}(t) \rangle = \int_0^t \bar{\epsilon}_i(s) r'_{i,j}(s)ds \\ 
    \langle \left( \Gamma_j^{(1)} \right)^2\rangle =  \sum_{ik}\langle R_{j}^{\epsilon(i)}(T) \rangle  \langle R_{j}^{\epsilon(k)}(T) \rangle\\ 
    + \sum_{ik}\int_{-\infty}^{\infty} S_{ik}(\omega) F_{ij}^{(1)*}(\omega,T)F_{kj}^{(1)}(\omega,T) d\omega\\ 
\end{split}
\end{equation}
here the 1st-order \emph{fundamental filter function} is defined as $F_{ij}^{(1)}(\omega,t)=\int_0^t C_\epsilon^{(i)}(s) r_{i,j}'(s)e^{i \omega s} ds$ and $C_\epsilon^{(i)}$ is a time-dependent factor that accounts for changes in noise amplitude over time. 
For example, in the case of multiplicative amplitude noise, we have $C_\epsilon(t)=\Omega(t)$, and $\epsilon_\Omega(t)=\tilde{\epsilon}(t) \Omega(t)$ where $\tilde{\epsilon}(t)$ is a stationary stochastic process. 
The power spectral density (PSD) $S_{ik}(\omega)=\frac{1}{2\pi}\int_{-\infty}^{\infty} ds e^{-i \omega s} \langle \tilde{\epsilon}_i(s) \tilde{\epsilon}_k(0)\rangle$. 
While this expression is general, many physical noise sources are either stationary with zero mean or effectively quasistatic. In such cases the expression simplifies considerably, e.g., in the quasistatic limit $\Gamma_j^{(1)}=\sum_i \epsilon_i r_{i,j}(T)$, and the QEED curve alone fully characterizes the first-order error.

Although the same approach can be applied to higher-order terms, their expressions can become more complex as high-order correlations are involved, and we will not discuss them here.

\section{Perturbative Expansion Of The Noise Channel}
\label{noisech}
In this section we derive the analytical formula for the noise channel when RC is applied to the circuit.
We have in general
\begin{equation}
\begin{split}
    \mathcal{E}(P)&=\exp{-i \Phi} P \exp{i \Phi}\\
    &\Phi=\sum_{n=1}^{\infty} \frac{1}{n!}\sum_i \epsilon^n \Gamma_i^{(n)} \sigma_i
\end{split}
\end{equation}
where $\Gamma_i^{(n)}$ is the n-th order error coefficient calculated through Magnus expansion. We can expand the expression using the Baker-Campbell-Hausdorff formula:
\begin{equation}
    \begin{split}
        \mathcal{E}(P)&=\sum_{n=0}^{\infty}\mathcal{E}_{\text{BCH}}^{n}(P)\\
        \mathcal{E}_{\text{BCH}}^{n}(P)&=\frac{(-i)^n}{n!}
        \sum_{\substack{i_1,i_2,...,i_n\\m_1,m_2,...,m_n}
                                    }\prod_{j=1}^n\frac{\Gamma_{i_j}^{(m_j)}}{m_j!}[\sigma_{i_1}, [\sigma_{i_2},...[\sigma_{i_n},P]]...]
    \end{split}
\end{equation}
However, for the purpose of characterizing the resulting noise channel, it is more useful to reorganize the BCH expansion according to powers of the noise strength $\epsilon$. We therefore write $\mathcal{E}(P)=\sum_n \mathcal{E}^{n}_\epsilon(P)$, where each term scales as $\epsilon^n$. Further, write the expansion to get: 
\begin{equation}
    \mathcal{E}^{n}_\epsilon(P) = \mathcal{E}^{n,1}_\epsilon(P) + \mathcal{E}^{n,2}_\epsilon(P)+...+\mathcal{E}^{n,n}_\epsilon(P)  
\end{equation}
whereas each $\mathcal{E}^{n,k}(P)$ is the contribution of n-th order term of the expanded noise channel contributed by BCH expansion $\mathcal{E}_{\text{BCH}}^{k}(P)$.

This can be further expressed as 
\begin{equation}
    \begin{split}
        &\mathcal{E}^{n,k}(P)=\frac{(-i)^k}{k!}\sum_{i_1,i_2,...,i_k}\mathcal{K}_{i_1,i_2,...i_k}^{n}\mathcal{P}_{i_1,i_2,...i_k}\\
        &\mathcal{K}_{i_1,i_2,...i_k}^{n} = \sum_{\sum_j m_j=n}\frac{\Gamma_{i_1}^{(m_1)} \Gamma_{i_2}^{(m_2)}... \Gamma_{i_k}^{(m_k)}}{m_1! m_2!...m_k!}\\
        &\mathcal{P}_{i_1,i_2,...i_k} = [\sigma_{i_2},...[\sigma_{i_n},P]]...]
    \end{split}
\end{equation}

It's easy to see that $\mathcal{K}$ has a symmetrical feature on all its indices: $\mathcal{K}_{...i_\alpha,...,i_\beta,...}^{n}=\mathcal{K}_{...i_\beta,...,i_\alpha,...}^{n}$. Meanwhile, for $\mathcal{P}$, reordering $i$'s will add a plus or minus sign before it, making $\mathcal{P}$ either symmetrical or anti-symmetrical about its indices.
As a result, the summation of $\mathcal{K}_{i_1,i_2,...i_k}^{n}\mathcal{P}_{i_1,i_2,...i_k}$ can be reduced to cases where $\mathcal{P}$ exhibits the same symmetry, as all combinations of $\sigma$'s that making $\mathcal{P}$ anti-symmetrical will vanish. 
This requires $[\sigma_{i_\alpha}, \sigma_{i_\beta}]=0$ for all pairs of indices $(i_\alpha, i_\beta)$ in $\mathcal{P}$. This also requires that all $\sigma_i$ in $\mathcal{P}$ anti-commute with $P$ as otherwise $\mathcal{P}$ becomes zero. In this case, $\mathcal{P}_{i_1,i_2,...,i_k}=2^k \sigma_{i_1}\sigma_{i_2}...\sigma_{i_k}P$.

Now consider cases when RC is applied. With random twirling gate $T$, all Pauli operators appears in $\mathcal{P}$ becomes $T^\dagger \sigma T$ and therefore we have $\tilde{\mathcal{P}}_{i_1,i_2,...i_k} = 2^k T^\dagger \sigma_{i_1}\sigma_{i_2}...\sigma_{i_n} T P$. 
After averaging over random twirling gates, we obtain $\mathbb{E}_T(T^\dagger \sigma_{i_1}\sigma_{i_2}...\sigma_{i_n} T)=0$ for $\sigma_{i_1}\sigma_{i_2}...\sigma_{i_n} \neq \mathcal{I}$. 
Consequently, for most of the index configurations of $\mathcal{P}$, we get a zero after performing the average, except for cases where the Pauli product chain forms an identity operator, giving another constraint on the combination of indices of $\mathcal{K}^{n}$ and $\mathcal{P}$. 
The resulting noise channel, after randomized compiling, is then transformed from a coherent noise channel to a Pauli channel, as we have $\mathcal{P}\propto P$. 

The constraints on the Pauli product chain also affect the number of $k$. 
Note that we always have $[\sigma_{i_1}\sigma_{i_2}...\sigma_{i_{k-1}}, P]=0$ for odd $k$, and $\sigma_{i_{k}}=\pm\sigma_{i_1}\sigma_{i_2}...\sigma_{i_{k-1}}$ must be satisfied for the entire product chain to equal the identity. 
This conflicts with the requirement that $\sigma_{i_{k}}$ shall anticommute with $P$, and therefore $k$ must be even.

\end{document}